%Paper: gr-qc/9303026
%From: Alan Rendall <RENDALL@sbitp.itp.ucsb.edu>
%Date: Sat, 20 Mar 1993 11:27 PST

\magnification=1200
\def\next{\hfil\break\noindent}
\def\ref#1{\lbrack #1\rbrack}
\def\ip{\langle\ ,\ \rangle}
\def\R{{\bf R}}
\def\C{{\bf C}}
\font\title=cmbx12
{\title \centerline{ Unique determination of an inner product by
adjointness relations}
\centerline{
in the algebra of quantum observables}}

\vskip 1cm

\next
Alan D. Rendall
\footnote*{Supported in part by the NSF grant PHY90-16733 and research
funds provided by Syracuse University}
\footnote{}{Permanent address: Max-Planck-Institut f\"ur
Astrophysik, Karl-Schwarzschild-Str. 1, 8046 Garching bei M\"unchen,
Germany.}
\next
Physics Department
\next
Syracuse University
\next
Syracuse NY 13244-1130
\next
USA

\vskip 1cm
\noindent
{\bf Abstract}

It is shown that if a representation of a *-algebra on a vector space
$V$ is an irreducible *-representation with respect to some inner
product on $V$ then under appropriate technical conditions this
property determines the inner product uniquely up to a constant
factor. Ashtekar has suggested using the condition that a given
representation of the algebra of quantum observables is a
*-representation to fix the inner product on the space of
physical states. This idea is of particular interest for
the quantisation of gravity where an obvious prescription for
defining an inner product is lacking. The results of this paper show
rigorously that Ashtekar's criterion does suffice to determine the inner
product in very general circumstances. Two versions of the result are
proved: a simpler one which only applies to representations by bounded
operators and a more general one which allows for unbounded operators.
Some concrete examples are worked out in order to illustrate the meaning
and range of applicability of the general theorems.

\vskip 1cm
\noindent
Short title: Determination of inner product by adjointness relations


\vfil\eject

\noindent
{\bf 1.Introduction}

There exist many approaches to the issue of giving a general definition
of what it means to quantise a classical system. Many of these involve
defining an abstract algebra which incorporates various algebraic
properties of classical observables and then studying representations
of this algebra on a Hilbert space. Recently Ashtekar (\ref1,
Chapter 10)  has put forward a quantisation programme which is designed
to be applicable to canonical quantum gravity. This has been
investigated further in \ref2. Since the Hamiltonian formulation
of general relativity involves constraints the programme must be able to
handle quantisation of constrained systems and in fact it is an
extension of the Dirac procedure for quantising such systems. However
the Dirac method does not give a prescription for finding an inner
product on the space of physical states. In many situations there is
some \lq background\rq\ structure which can be used to fix this inner
product. For instance, in the case of a field theory in flat space there
is the background Minkowski metric and the associated notion of
Poincar\'e invariance can be used to restrict the inner product. On the
other hand the absence of any background structure is a characteristic
feature of general relativity. To overcome this problem, Ashtekar gives
a general criterion for choosing an inner product which is the
following. Suppose that for a given classical system we have defined an
abstract algebra $A$ (the algebra of quantum observables) and that by
some means we have found a representation of $A$ on a complex vector
space $V$. In general there is at this stage no preferred inner product
on $V$. Suppose further that the structure of the classical system gives
rise to a natural involution on the algebra, $a\mapsto a^*$. Then the
inner product is to be restricted by the requirement that this
involution goes over into the operation of taking the
adjoint of an operator. To be more precise, any inner product on $V$
gives rise to a definition of the adjoint of a linear operator defined
on $V$. If $\rho$ denotes the representation of $A$ on $V$ then for any
$a\in A$ the adjoint of the linear operator $\rho(a)$ is required to be
equal to $\rho(a^*)$.

The purpose of this paper is to show that the above proposal determines
the inner product up to a constant factor under rather general
circumstances. The first task which has to be accomplished in doing
this is to find an appropriate mathematical formulation of the problem.
There are two constraints involved in doing this. On the one hand, the
desired result will fail if the technical hypotheses are not strong
enough. On the other hand, the result will only be of interest from the
point of  view of the original motivation if the hypotheses hold (and
can reasonably be verified) for a large class of systems which one might
wish to quantise. The ambiguities in the mathematical formulation will
now be discussed. The first point is that in the end the representation
of the algebra $A$ should be a representation on a Hilbert space. In
general this Hilbert space will not be $V$ but rather the completion of
$V$ with respect to the inner product which is to be determined.
Moreover, the elements of $A$ may be represented by unbounded operators
on this Hilbert space which leads, as is usual with unbounded operators,
to technical difficulties with the domains of definition of the
operators. For this reason, the first theorem in this paper will
be concerned only with the case where all the elements of $A$ can be
represented by bounded operators. This theorem will frequently {\it not}
be applicable to the algebra of quantum operators in the form in which
it most naturally arises in a given physical problem. The second theorem
does allow unbounded operators but requires more technical assumptions.
These technical assumptions are such that it should often be practicable
to check them in interesting examples. Nevertheless it may be easier to
change the algebra of quantum observables itself so as to allow the
theorem on representations by bounded operators to be applied. The
standard example of such a modification of the algebra of
observables for mathematical convenience is the replacement of the
algebra of the canonical commutation relations by the Weyl algebra (the
\lq integrated form\rq\ of the commutation relations).

Even when only bounded operators are considered, there is another
potential ambiguity. The inner product can only be expected to be unique
if some assumption of irreducibility is made on the representation.
Without this it would be possible to choose the inner product
independently in different superselection sectors. There are two
candidates for the definition of irreducibility. These are {\it
algebraic} and {\it topological} irreducibility. A representation of an
algebra $A$ on a vector space is algebraically irreducible if there are
no invariant subspaces; if the vector space has a topology (e.g. if it
is a Hilbert space) then the representation is said to be topologically
irreducible if there are no {\it closed} invariant subspaces. The first
of these definitions seems attractive in the context of the present
application since it only involves the vector space structure
of $V$ while to define topological irreducibility it is necessary to
make use of the inner product to be determined. However it turns out
that algebraic irreducibility is an unreasonably strong assumption. To
see this, consider the standard representation of the Weyl relations for
one degree of freedom on $L^2(\R)$. Thus $U(a)$ is represented by
translation by $a$ and $V(b)$ is represented by multiplication by the
function $e^{ibq}$. Examples of invariant subspaces are the Schwartz
space of rapidly decreasing functions, the smooth functions of compact
support and the space of functions whose Fourier transforms are smooth
and have compact support. For this reason only topological
irreducibility is considered in the following.

The paper is organised as follows. In section 2 a theorem on the
uniqueness
of inner products is proved in the case that the elements of $A$ can be
represented by bounded operators. Section 3 contains a generalisation of
this to unbounded operators whose proof makes use of the previous
result.
Some illustrative examples are presented in the final section.

\vskip .5cm
\noindent
{\bf 2. The case of bounded operators}

Recall that a complex *-algebra is an associative algebra over $\C$
together with a mapping $a\mapsto a^*$ which is antilinear and satisfies
the conditions that $(ab)^*=b^*a^*$ for any $a,b\in A$ and that
$(a^*)^*=a$ for any $a\in A$.

\noindent
{\bf Definition 1} Let $A$ be a complex *-algebra with identity and
$\rho$ a representation of $A$ on a complex vector space $V$.
An inner product $\ip_1$ on $V$ is said to be
{\it strongly admissible} if:
\next
(i) $\rho$ is a *-representation with respect to this inner product
i.e. $\langle\rho(a)x,y\rangle_1=\langle x,\rho(a^*)y\rangle_1$
for all $a\in A$ and $x,y\in V$
\next
(ii) for each $a\in A$ the operator $\rho(a)$ is bounded with respect
to the norm $\|\ \|_1$ defined by the given inner product so that $\rho$
extends uniquely to a representation $\hat\rho_1$ on the Hilbert space
completion $\hat V_1$ of $V$ by bounded operators
\next
(iii) $\hat\rho_1$ is topologically irreducible.
\vskip .25cm
\noindent
{\bf Theorem 1} Let $\ip_1$ and $\ip_2$ be inner products on the complex
vector space $V$ which are strongly admissible with respect to a
representation $\rho$ of a complex *-algebra $A$. Then $\ip_1=c\ip_2$
for some positive real number $c$.

\noindent
{\bf Proof} Suppose that $\ip_1$ and $\ip_2$ are strongly admissible
inner products. Define
$$\ip=\ip_1+\ip_2.\eqno(1)$$
This defines a Hermitian form on $V$. Now (1) implies that $\langle x,
x\rangle\ge0$ for all $x$ and moreover $\langle x,x\rangle=0$
implies that $\|x\|_1=0$ and hence that $x=0$. Hence $\ip$ defines an
inner product and the associated norm satisfies
$$\eqalign{
&\|x\|_1\le\|x\|,              \cr
&\|x\|_2\le\|x\|,}\eqno(2)$$
for all $x\in V$. Let $\hat V$ be the Hilbert space completion of $V$
with respect to $\ip$. The boundedness of the operators $\rho(a)$
with respect to $\ip_1$ and $\ip_2$ implies that they are bounded
with respect to $\ip$. Hence $\rho$ extends uniquely to a representation
$\hat\rho$ on $\hat V$ by bounded operators. Now
$$\langle x,y\rangle_1\le \|x\|_1\|y\|_1\le\|x\|\|y\|\eqno(3)$$
for all $x$, $y$ in $V$. Thus $\ip_1$ extends uniquely to a continuous
Hermitian form on $\hat V$. The extension, which will also be denoted
by $\ip_1$, still satisfies (3). By continuity the relation
$$\langle\rho(a)x,y\rangle_1=\langle x,\rho(a^*)y\rangle_1\eqno(4)$$
continues to hold for the extension. Consider now the mapping
$x\mapsto \langle x,y\rangle_1$. As a result of (3) this is a continuous
linear functional on $\hat V$. By Riesz' lemma there exists a
$z\in\hat V$ with $\langle x,y\rangle_1=\langle x,z\rangle$. Doing this
for each $y$ gives a linear operator $L_1:y\mapsto z$ which is bounded
by (3). This operator satisfies $\langle x,y\rangle_1=\langle x,L_1 y
\rangle$ and is self-adjoint. Now
$$\langle x,L_1\rho(a)y\rangle=\langle x,\rho(a)L_1y\rangle,\eqno(5)$$
for all $x,y\in\hat V$. Here the fact has been used that $\ip$ satisfies
the condition (i) of a strongly admissible inner product, which follows
directly from its definition. Hence $L_1$ commutes with $\rho(a)$ for
all $a\in A$. Since $\|x\|_1\le\|x\|$ the identity mapping of $V$
extends uniquely to a continuous linear mapping $Q:\hat V\to\hat V_1$.
It has already been mentioned that the representation $\rho$ extends to
both $\hat V$ and $\hat V_1$ in a natural way. Given $x\in\hat V$ choose
a sequence $x_n$ in $V$ such that $\|x-x_n\|\to 0$. Then $\hat\rho(a)$
is defined by the condition that $\|\hat\rho(a)x-\rho(a)x_n\|\to 0$. Now
$\|Q(x)-x_n\|_1\to 0$ and so $\|\hat\rho_1(a)Q(x)-\rho(a)x_n\|_1\to0$.
This implies that $\hat\rho_1(a)\circ Q=Q\circ\hat\rho(a)$.

The operator $L_1$ is positive, bounded and self-adjoint and so its
spectrum must be contained in the interval $[0,\lambda_1]$ for some
$\lambda_1>0$. Choose $\lambda_1$ to be minimal for this property.
There are now two cases to be considered. Case (i) is that the spectrum
of $L_1$ is equal to $\{\lambda_1\}$. Then $L_1=\lambda_1 I$ and
$\ip_1=\lambda_1\ip$. In case (ii) the infimum $\lambda_1^\prime$ of
the numbers in the spectrum of $L_1$ is strictly less than $\lambda_1$.
Then there exists some $\mu$ with $\lambda^\prime_1<\mu<\lambda_1$.
Let $\Pi=\int_\mu^{\lambda_1} dE(\lambda)$, where $dE(\lambda)$ is the
spectral measure of $L_1$. (For information concerning spectral measures
see \ref4\ , chapters 12 and 13.) This projection is not equal to zero
or the identity. Hence its image, $K$ say, is a closed linear subspace
of $\hat V$ different from $\{0\}$ and $\hat V$. Since all the $\rho(a)$
commute with $L_1$ they also commute with $\Pi$ and hence $K$ is an
invariant closed linear subspace for the representation $\hat\rho$. If
it was known that $\hat\rho$ was irreducible this would give a
contradiction and complete the proof. However it is not clear that the
irreducibility of $\hat\rho_1$ and $\hat\rho_2$ implies that of
$\hat\rho$ and a less direct approach will be adopted.

Now $(L-\mu I)|_K$ is an operator with positive spectrum and hence a
positive operator. Thus $\langle L_1x,
x\rangle\ge\mu\langle x,x\rangle$ or $\|x\|\le\mu^{-1/2}\|x\|_1$ on $K$.
It follows that the two norms $\|\ \|$ and $\|\ \|_1$ are equivalent
when restricted to $K$. This implies in particular that the restriction
of $Q$ to $K$ is injective so that $K$ can be regarded as a linear
subspace of $\hat V_1$. The subspace $K$ is closed with respect to
$\|\ \|$ and hence complete. It follows that it is complete with respect
to $\|\ \|_1$ and hence closed in $\hat V_1$. It is also invariant under
$\hat\rho_1$. Since $\hat\rho_1$ is irreducible this is impossible
unless $K=\hat V_1$. Suppose that, for a given choice of $\mu$, $K=\hat
V_1$. Let $\tilde\mu$ be a number in the interval $(0,\mu)$, let
$\tilde \Pi=\int_{\tilde\mu}^{\lambda_1} dE(\lambda)$ and let $\tilde K$
be the image of $\tilde\Pi$. By the same argument as before $Q|\tilde K$
is injective and so $K=\hat V_1$ implies that $K=\tilde K$. If
$\lambda_1^\prime$ were greater than zero then choosing
$\tilde\mu<\lambda_1^\prime$ would give a contradiction to the fact that
$K$ is not the whole of $\hat V$. Hence $\lambda_1^\prime=0$ and
moreover, for all $\epsilon$ sufficiently small,
$\int_0^\epsilon dE(\lambda)$ is the orthogonal projection on the
orthogonal complement of $K$. However this means that the value of the
spectral measure on the set $\{0\}$ is also this projection. It can be
concluded that the orthogonal complement of $K$ is the kernel of $L_1$.
If $x\in ker L_1$ then $\langle x,x\rangle_1=0$. The non-degeneracy of
$\ip_1$ thus implies that the kernel of $L_1$ is trivial. This shows
that the assumption that case (ii) holds necessarily leads to a
contradiction. It follows that only case (i) is in fact realised. The
conclusion of all this is that $\ip_1=\lambda_1\ip$.
Similarly $\ip_2=\lambda_2\ip$ for some $\lambda_2>0$.
So $\ip_1=(\lambda_1/\lambda_2)\ip_2$.

\vskip .5cm\noindent
{\bf 3. The case of unbounded operators}

For applications it is useful to have a version of Theorem 1
where the assumption (ii) in the definition of a strongly
admissible inner product is weakened to allow representations by
unbounded operators. If we have a representation $\rho$ of an algebra
$A$ on a vector space $V$ as before and an inner product $\ip$ on $V$
then $\rho$ can be regarded as a representation of $A$ on $\hat V$ by
unbounded operators with domain $V$. For the sake of clarity this
representation by unbounded operators will be denoted by $\hat\rho$.
The assumptions for a generalisation of Theorem 1 should include the
irreducibility of this representation and so it is necessary at this
point to introduce a definition of irreducibility of a representation
by unbounded operators. In fact there is more than one way in which
one might think of defining irreducibility for a representation by
unbounded operators. The one used here agrees with that used in \ref5.
If two representations $\rho_1$ and $\rho_2$ are given on Hilbert spaces
$H_1$ and $H_2$ with domains $D_1$ and $D_2$ respectively, then
a representation on $H_1\oplus H_2$ with domain $D_1\oplus D_2$
is obtained by defining $(\rho_1\oplus\rho_2)(a)(x_1,x_2)=
(\rho_1(a)x_1,\rho_2(a)x_2)$. The representation $\rho_1\oplus
\rho_2$ is called the direct sum of $\rho_1$ and $\rho_2$. A
representation is called irreducible if it cannot be written in
a non-trivial way as a direct sum of representations. Note that in
the case of a representation by bounded operators this definition
agrees with the notion of topological irreducibility used in the
previous section.

Another concept which will be required in the following
is that of a closed representation. It is a generalisation of the
concept of a closed operator. If $L$ is an operator on the Hilbert space
$H$ with domain $D$ then its {\it graph} is the subset of $H\times H$
consisting of pairs of the form $(x,Lx)$ with $x\in D$. The operator is
called {\it closed} if its graph is a closed subset of $H\times H$. More
generally it may be that if the graph of an operator is not closed the
closure of its graph is the graph of an operator $\bar L$. This operator
is uniquely determined and is called the closure of $L$.
It is an extension of $L$. Now suppose that $A$ is a *-algebra and
$\rho$ a *-representation of $A$ on a Hilbert space $H$ with domain $D$.
It can be shown that each operator $\rho(a)$ has a closure. Let $\bar D$
be the intersection of the domains of the closures of the operators
$\rho(a)$ as $a$ runs over $A$. This is a subspace of $H$ containing
$D$. If in fact $D=\bar D$ then the representation is called closed.
Further details can be found in \ref5.

\noindent
{\bf Definition 2} Let $A$, $\rho$ and $V$ be as in Definition 1.
Let $S$ be a set of elements of $A$ which satisfy $a^*=a$ and
which generate $A$. An inner product $\ip_1$ on is said to be
{\it admissible} if:
\next
(i) $\rho$ is a *-representation with respect to this inner product
\next
(ii) for each $a\in S$ the operator $\hat\rho_1(a)$ is essentially
self-adjoint
\next
(iii) $\hat\rho_1$ is irreducible
\next
(iv) $\hat\rho_1$ is closed

\vskip .25cm\noindent
{\bf Remarks} 1. The algebra generated by a set of elements of a given
algebra means here and in the following the smallest subalgebra
containing these elements and the identity.
\next
2.As implied by the terminology, a representation which
is strongly admissible is admissible. For in that case $S$ can be taken
to be the set of all $a\in A$ satisfying $a^*=a$.

\vskip .25cm\noindent
{\bf Theorem 2} Let $\ip_1$ and $\ip_2$ be inner products on the complex
vector space $V$ which are admissible with respect to a representation
$\rho$ of a complex *-algebra $A$. Then $\ip_1=c\ip_2$ for some positive
real number $c$.
\next
{\bf Proof} If $a\in S$ define $F_1(a)$ to be the closure of
$\hat\rho_1(a)$. By assumption $F_1(a)$ is self-adjoint and so it is
possible to define $G_1(a)=(i+F_1(a))/(i-F_1(a))$. The operator $G_1(a)$
satisfies the equation
$$(i-F_1(a))G_1(a)=i+F_1(a).\eqno(6)$$
An operator $G_2(a)$ can be defined similarly using the inner product
$\ip_2$. Now the restrictions of $F_1(a)$ and $F_2(a)$ to $V$ are both
equal to $\rho(a)$. Hence (6) and the analogous equation satisfied by
$F_2(a)$ and $G_2(a)$ imply that $(i-\rho(a))(G_1(a)-G_2(a))|_V=0$.
Since $\rho(a)$ is essentially self-adjoint its deficiency indices must
be zero and so $i-\rho(a)$ has trivial kernel. It follows that the
restrictions of  $G_1(a)$ and $G_2(a)$ to $V$ are equal. Let
$b(a)=G_1(a)|_V$. Now define a new algebra $B$ to consist of the algebra
of linear mappings generated by the $b(a)$ as $a$ runs through the set
$S$. Every element of $B$ possesses an adjoint in the algebra and this
defines a star operation in $B$. Consider now the defining
representation of $B$ on $V$. It extends to *-representations of $B$ on
$\hat V_1$ and $\hat V_2$ by bounded operators. Thus the representation
of $B$ satisfies conditions (i) and (ii) of Definition 1.

To examine whether it also satisfies condition (iii), suppose that $K$
is a closed linear subspace left invariant by the extensions to $\hat
V_1$ of all elements of $B$. In particular, if $a\in S$ then $G_1(a)$
commutes with the projection $\pi_K$ on $K$. This means that $\pi_K$
commutes with all spectral projections of $G_1(a)$, which in turn means
that it commutes with all spectral projections of $F_1(a)$. It follows
that $\pi_K$ maps the domain of $F_1(a)$ into itself and that for any
$x$ in this domain $\pi_KF_1(a)x=F_1(a)\pi_Kx$.  Hence condition (iv)
implies that $\pi_K$ maps $V$ into itself. Since elements of $S$
generate $A$ it can be concluded that $\pi_K\rho(a)x=\rho(a)\pi_Kx$ for
all $a\in A$ and $x\in V$. Using the fact that $\hat\rho_1$ is
irreducible and applying Lemma 8.3.5 of \ref5\ shows that $K$ must be
the trivial subspace or the whole of $\hat V_1$. It has now been shown
that $\ip_1$ is strongly admissible for the representation of $B$ and of
course the same argument can be applied to $\ip_2$. With this
information the conclusion of Theorem 2 for the algebra $A$ is obtained
by applying Theorem 1 to $B$.

\vskip .5cm\noindent
{\bf 4. Examples}

The first example to be considered is one which is rather simple from
the point of view of the general theory developed above.
This is the case where the vector space $V$ is finite dimensional. In
this case there are several simplifications. When $V$ is finite
dimensional it is complete with respect to any inner product so that
$V=\hat V_1=\hat V_2$. Moreover all linear operators are bounded and
so in this case Theorem 2 is unnecessary. Only properties (i) and (iii)
need to be checked in order to apply Theorem 1. The distinction between
algebraic and topological irreducibility also becomes superfluous since
in finite dimensions all linear subspaces are closed. Finally, it should
be remarked that Theorem 1 is much easier to prove when $V$ is finite
dimensional; the result is then a rather straightforward application of
the classical Schur lemma. One of the constrained systems whose
quantisation is discussed in \ref2\ consists of two harmonic oscillators
which are required to have a fixed total energy. This is a situation
where Theorem 1 can be applied to a finite dimensional representation.

The remaining examples are intended to illustrate how the hypotheses of
the above theorems can be checked in practice. They also give some idea
of the relative merits of Theorems 1 and 2 in applications. The first is
the Weyl algebra for one degree of freedom and the second the algebra of
the canonical commutation relations. Of course these are just two
alternative algebras of quantum operators for a single classical system.
Then the case of more (possibly infinitely many) degrees of freedom is
examined.

The Weyl algebra for one degree of freedom is the algebra generated by
elements $U(a)$ and $V(b)$ with $a,b\in\R$ subject to the relations
$$\eqalign{
&U(a+b)=U(a)U(b),\ \ \ V(a+b)=V(a)V(b),                            \cr
&U(a)V(b)=e^{iab}V(b)U(a).}\eqno(7)$$
The star relations are $U(a)^*=U(-a)$ and $V(b)^*=V(-b)$. The standard
representation is that on $L^2(\R)$ given by taking $U(a)$ to be
translation by $a$ and $V(b)$ to be multiplication by $e^{ibq}$. It
is well known that this representation is irreducible and in fact it
is the unique weakly continuous irreducible representation up to unitary
equivalence (von Neumann uniqueness theorem). In order to demonstrate
what is needed to check the hypotheses of Theorem 1 a proof of the
irreducibility will be now be given. (The other hypotheses are easily
verified.) The method of proof is useful in more general situations. To
show that the representation is irreducible it is sufficient to show
that any projection $\Pi$ which commutes with all $U(a)$ and $V(b)$ is
either equal to zero or the identity. Suppose then that $\Pi$ is a
projection which commutes with all elements of the Weyl algebra. It is
useful at this point to bear in mind the following vague but helpful
principle: any operator on $L^2(X)$, where $X$ is some measure space,
which commutes with sufficiently many multiplication operators is itself
a multiplication operator. Denote by $M_f$ the multiplication operator
associated to $f$ i.e. $M_f(g)=fg$. If $f,g$ are two smooth functions of
compact support on $\R$, choose $R$ so that the supports of both are
contained in the interval $\lbrack -R, R\rbrack$. The function $f$ can
be altered outside this interval to give a function $f_R$ which is
periodic with period $2R$. This can be uniformly approximated by a
sequence of functions $f_{R,n}$, each of which is a finite linear
combination of the functions $e^{ibq}$. For instance this could be the
sequence of partial sums of the Fourier series of $f_R$. Now the
projection $\Pi$ commutes with $M_{f_{R,n}}$ for each $n$. Moreover
$M_{f_{R,n}}$ converges to $M_{f_R}$ in the norm topology as
$n\to\infty$. Thus
$$M_{f_R}\Pi g=\Pi M_{f_R}g=\Pi M_fg.$$
If $R$ is allowed to tend to infinity then $M_{f_R}(\Pi g)$ will
converge to $M_f(\Pi g)$ in $L^2(\R)$.
It follows that $M_f\Pi g=\Pi M_fg$. Since $g$ was
an arbitrary smooth function of compact support this implies that $M_f$
commutes with $\Pi$. Now let $h=e^{x^2}\Pi (e^{-x^2})$. Then
$$\eqalign{M_h(e^{-x^2}f)&=e^{-x^2}fe^{x^2}\Pi (e^{-x^2})
                         =M_f\Pi (e^{-x^2})               \cr
                        &=\Pi M_f(e^{-x^2})
                         =\Pi (e^{-x^2}f)}$$
Since this is true for all smooth $f$ of compact support it follows
that $\Pi=M_h$. A multiplication operator which is a projection must
be that associated with the characteristic function of some measurable
set, $h=\chi_E$. But now, using the fact that $\Pi$ commutes with all
$U(a)$, we see that the set $E$ is invariant under all translations.
This means that $\Pi$ can only be zero or the identity and the proof of
irreducibility is complete.

Next the canonical commutation relations for one degree of freedom will
be examined. Here the algebra is generated by two elements $Q,P$
satisfying $\lbrack Q,P\rbrack =i$. The star relations are $Q^*=Q$ and
$P^*=P$. In the standard Schr\"odinger representation $Q$ is represented
by multiplication by $q$ and $P$ by $-id/dq$, considered as unbounded
operators on $L^2(\R)$. It will be shown that, if a common domain for
these operators is chosen appropriately, the inner product of $L^2(\R)$
is admissible with respect to this representation. A dense subspace of
$L^2(\R)$ where the operators representing all elements of the algebra
are defined is $V_0=C_c^\infty(\R)$, the smooth functions of compact
support. To check condition (iv) in the definition of an admissible
inner product it is necessary to find a domain for which the
representation is closed. It will be shown that the Schwartz space
$\cal S$ is such a domain. If $L$ is an operator on a Hilbert space $H$
with domain $D$ the domain of its closure consists of all elements $f$
of $H$ with the following property: there exists an element $g\in H$
and a sequence $\{f_n\}$ in $D$ such that $f_n\to f$ and $Lf_n\to g$.
If $f\in\cal S$ there exists a sequence $\{f_n\}$ in $V_0$ such that
$f_n\to f$ in the topology of $\cal S$. In particular, if $L=Q^kP^l$
then $Q^kP^lf_n\to Q^lP^lf$ in $L^2(\R)$ so that $f$ belongs to the
closure of $L$. Conversely, suppose that $f$ belongs to the closure of
the operator $Q^kP^l$ with domain $V_0$ for all $k,l$. Thus for any
given $k,l$ there exists a sequence $\{f_n\}$ in $V_0$ such that
$f_n\to f$ and $q^kd^lf_n/dq^l\to g$ in $L^2(\R)$. From the first of
these properties of the sequence $\{f_n\}$ it can be concluded that
$q^kd^lf_n/dq^l\to q^kd^lf/dq^l$ in the sense of distributions. Thus
$g=q^kd^lf/dq^l$, where $d^lf/dq^l$ is to be interpreted as a
distributional derivative. It can now be seen that for fixed $k,l$
all derivatives of $q^kd^lf/dq^l$ belong to $L^2(\R)$. By the Sobolev
embedding theorem it follows that $q^kd^lf/dq^l$ is continuous and
bounded. Hence $f$ is in $\cal S$.

Take $V=\cal S$. Then properties (i) and (iv) of an admissible inner
product are satisfied. To verify the other properties choose
$S=\{Q,P\}$. The essential self-adjointness of $Q$ and $P$ can be
checked by looking at their deficiency indices. The equations $Qf=\pm
if$ and $Pf=\pm if$ have no solutions in $L^2(\R)$. This means that the
deficiency indices of $Q$ and $P$ are zero so that they are indeed
essentially self-adjoint. It remains to show that the representation is
irreducible. This means showing that the only projections $\Pi$ which
map $\cal S$ into itself and which commute with $Q$ and $P$ on that
domain are zero and the identity. If $\Pi$ commutes with $Q$ then it
commutes with any spectral projection of $Q$ (\ref4, Theorem 13.33).
This in turn implies that $\Pi$ commutes with every bounded measurable
function of $Q$ (\ref4, Theorem 12.21). In other words, it
commutes with the multiplication operators $M_f$ for any $f\in
L^\infty(\R)$. It has been shown above that any projection which
commutes with $M_f$ for all $f$ in the more restricted class of smooth
functions of compact support must be the multiplication operator defined
by the characteristic function of a measurable set $E$. Thus $\Pi=M_h$
where $h$ is of the form $\chi_E$ for some measurable set $E$.

The operator $\Pi$ maps $\cal S$ into itself. It will now be shown that
this can only be true if $E=\emptyset$ or $E=\R$. For otherwise the set
$E$ must have a boundary point $q_0$. Let $f$ be a function in the
Schwartz space which has the value one in a neighbourhood $U$ of $q_0$.
On that neighbourhood $\Pi f=\chi_E$ and since $q_0$ is a boundary point
of $E$ this means that $\Pi f$, like $\chi_E$, must be discontinuous
there. Hence $\Pi f$ is not in the Schwartz space. It can now be
concluded that $\Pi$ is zero or the identity and that the representation
is irreducible. It follows that the inner product of $L^2(\R)$ is
admissible with respect to the Schr\"odinger representation of the
algebra of the canonical commutation relations for one degree of
freedom.

The above arguments extend without difficulty to show that the inner
product of $L^2(\R^n)$ is strongly admissible with respect to the
Schr\"odinger representation of the Weyl algebra for $n$ degrees of
freedom and admissible for the corresponding representation of the
canonical commutation relations. In the case of infinitely many degrees
of freedom there still exists a Schr\"odinger representation of the
Weyl algebra (\ref3,\ section 6.4). However there are many other
unitarily inequivalent weakly continuous irreducible representations. It
should be stressed that the question of interest in this paper, namely
the unique determination of an inner product by a given representation,
is logically independent of the question of whether the representation
is unique under certain hypotheses. It is assumed throughout this paper
that not only the algebra $A$ but also the vector space $V$ and the
representation $\rho$ are given. The results then show that there is
at most one admissible inner product for the given representation.
No statement is made concerning the question whether, for a given
representation, any admissible inner product exists.

It turns out that Theorem 1 can be applied to the Schr\"odinger
representation of the Weyl algebra in the case of infinitely many
degrees of freedom. The above arguments for irreducibility do not
obviously generalise to the infinite dimensional case since techniques
have been used (Fourier series and distribution theory) which are
usually only considered for finite dimensional spaces. However there
exist other arguments to prove the irreducibility of the Schr\"odinger
representation for infinitely many degrees of freedom. It is, for
instance, possible to use the equivalence of this representation with
the Fock representation (\ref3, section 7.3). It follows that the
hypotheses of Theorem 1 are still satisfied in the infinite dimensional
case.

\next
{\it Acknowledgements}

\noindent
I am grateful to Abhay Ashtekar and Jorma Louko for helpful discussions.

\vskip .5cm\noindent
{\bf References}
\next
1. Ashtekar, A. (1991) Lectures on non-perturbative canonical gravity.
(World Scientific: Singapore)
\next
2. Ashtekar, A., Tate, R.S. An extension of the Dirac program for the
quantisation of constrained systems. Int. J. Mod. Phys. D (to appear)
\next
3. Bogolubov, N.N., Logunov, A.A., Oksak, A.I., Todorov, I.T. (1990)
General principles of quantum field theory. (Kluwer:Dordrecht)
\next
4. Rudin, W. (1991) Functional analysis. 2nd Edition. (McGraw-Hill:
New York)
\next
5. Schm\"udgen, K. (1990) Unbounded operator algebras and
representation theory. (Basel: Birkh\"auser)

\end